\documentclass[prd,aps,showpacs,nofootinbib,nobibnotes]{revtex4}

\usepackage{graphicx}

\usepackage{epsfig,psfrag}

\def\gsim{\;\rlap{\lower 2.5pt

 \hbox{$\sim$}}\raise 1.5pt\hbox{$>$}\;}

\def\lsim{\;\rlap{\lower 2.5pt

   \hbox{$\sim$}}\raise 1.5pt\hbox{$<$}\;}

\begin{document}

\title{Systematic Errors in Sunyaev-Zeldovich Surveys of Galaxy Cluster Velocities}
\author{Suman Bhattacharya}
\email{sub5@pitt.edu}
\author{Arthur Kosowsky}
\email{kosowsky@pitt.edu}
\affiliation{Department of Physics and Astronomy, University of Pittsburgh, 3941 O'Hara Street, Pittsburgh, PA 15260 USA}

\begin{abstract}

Galaxy cluster surveys compiled via the Sunyaev-Zeldovich Effect have the potential
to place strong constraints on cosmology, and in particular the nature of dark energy.
Here we consider cluster velocity surveys using
kinetic Sunyaev-Zeldovich measurements. Cluster velocities closely trace the large-scale
velocity field independent of cluster mass; we demonstrate that two useful cluster velocity
statistics are nearly independent of cluster mass, in marked
contrast to cluster number count statistics. On the other hand, cluster velocity determinations 
from three-band observations of Sunyaev-Zeldovich distortions can 
require additional cluster data or assumptions,
and are complicated by microwave emission from dusty galaxies and radio sources,
which may be
correlated with clusters. Systematic errors in velocity due to these factors can give substantial biases
in determination of dark energy parameters, although large cluster velocity surveys will contain enough 
information that the errors can be modeled using the data itself, with little degradation in
cosmological constraints. 

\end{abstract}


\maketitle

\section{Introduction}
\label{sec:intro}

The formation of galaxy clusters is a sensitive tracer of structure growth in the universe.
It is widely appreciated that the number of clusters larger than a given mass as a function
of redshift has the potential to place strong constraints on dark energy properties: cluster
number counts are one of four techniques considered in detail by the Dark Energy
Task Force. Currently operating Sunyaev-Zeldovich experiments \cite{kosowsky06,ruhl05}
will soon produce large catalogs of galaxy clusters to all redshifts, potentially fulfilling
this promise. But the sensitivity of cluster number counts as a probe of cosmology also
means that systematic errors in these number counts can substantially bias any cosmological
constraints \cite{francis05,holder01,haiman01,mol02,hut04}. 
Including information on the spatial distribution
of galaxy clustering, termed ``self-calibration,'' can help characterize and compensate
for systematic errors in number counts \cite{mm03, lima04, lima05}, but these techniques require
demanding further observations and rely on specific simplifying assumptions about cluster properties. 

We have recently explored an alternative use of galaxy clusters to constrain dark energy, via their
peculiar velocities \cite{peel02,bk06}. 
The kinematic Sunyaev-Zeldovich Effect measures an effective Doppler shift of clusters,
proportional to both the cluster line-of-sight peculiar velocity and its optical depth, so measurement of
this small blackbody distortion in the microwave background radiation holds the promise of directly measuring cluster velocities with respect to the rest frame of the microwave radiation. Cluster velocities
trace the large-scale velocity field arising from structure formation in the universe, and their velocities
are expected to be only weakly dependent on cluster mass. Therefore we expect that
dark energy constraints based on cluster velocities will be far less sensitive to systematic
errors in estimating the mass limit of any particular cluster catalog. We verify this expectation
here. While changing a cluster catalog mass cutoff by 20\% can change the total number
of clusters by a factor of two, it only changes cluster velocity statistics by a few percent. 
The bias on cosmological parameters from uncertainties in cluster selection will
be much milder for cluster velocity statistics than cluster number counts, and the
power of velocities to constrain dark energy is significant, even for modest velocity
errors as large as 500 km/sec \cite{bk07}. 

The kinematic SZ signal measures the total cluster baryon momentum, not the
cluster velocities directly. Extracting the velocities from SZ measurements will require
additional data to estimate the cluster baryon mass \cite{sehgal05}; cluster X-ray temperatures
are one likely route, while a suspected tight correlation between thermal SZ flux and gas temperature
is another. But these measurements have potential systematic errors of their own;
previous studies \cite{diaferio05, knox04} have shown that X-ray temperature systematically overestimates electron temperature by $20\%$ to $40\%$ (this particular systematic 
difference arises because while X-ray temperature is luminosity weighted, electron temperature is mass 
weighted), while correlations seen in numerical simulations may not incorporate all
of the relevant physical effects in real galaxy clusters. 
Observing the SZ signal will also be complicated by point source contamination,
which may induce a different systematic error. Using simple models for these errors, we show that
they give potentially significant biases to cosmological parameters if not properly accounted for.

In this work, we focus on two particular galaxy cluster velocity statistics: the correlation function of
the velocity components perpendicular to the line connecting a cluster pair,
$\left\langle v_iv_j\right\rangle_\perp(r)$, and the mean pairwise streaming velocity $v_{ij}(r)$; each is
a function of the separation between two galaxy clusters $r$ and redshift $z$. 
Both of these statistics were considered as probes of dark energy in Ref.~\cite{bk07}. (We drop the
perpendicular subscript from the correlation function for convenience). 
Sections \ref{sec:mass} and \ref{sec:xray} consider the systematic errors arising from uncertainty in the 
cluster mass selection function, and from systematic errors in velocity estimates due to misestimates in
galaxy cluster physics or contamination by foreground emission. Section \ref{sec:bias} then computes 
the resulting biases in
determining dark energy parameters for each source of error, while Section \ref{sec:selfcal} briefly 
considers self-calibration techniques in the context of cluster velocities.

\section{Systematic Errors from Mass Misestimates}  
\label{sec:mass}

In this section, we address the issue of mass selection in the context of cluster velocities.
The velocity statistics depend on $M_{\rm min}$ through the normalization 
term in theoretical halo models.  
In order to model the effect of mass selection, we assume that for a given survey, cluster masses
are all mis-estimated by a constant fraction. This leads to a corresponding difference in the inferred
cluster mass threshold for number statistics and the resulting systematic bias in dark energy parameters 
studied in Ref.~\cite{francis05}.  However, clusters of any mass generally trace the large-scale
velocity field, so biased cluster mass estimates should have little effect on cluster velocity
statistics. The two statistics $\left\langle v_iv_j\right\rangle_\perp(r)$ and $v_{ij}(r)$ can both
be estimated accurately with analytic approximations based on the halo model and on
nonlinear perturbation theory \cite{sheth01, sheth02, sheth04}; a summary of these approximations is 
given in Ref.~\cite{bk07}. 

\begin{figure*}
  \begin{center}
    \begin{tabular}{cc}
      \resizebox{85mm}{!}{\includegraphics{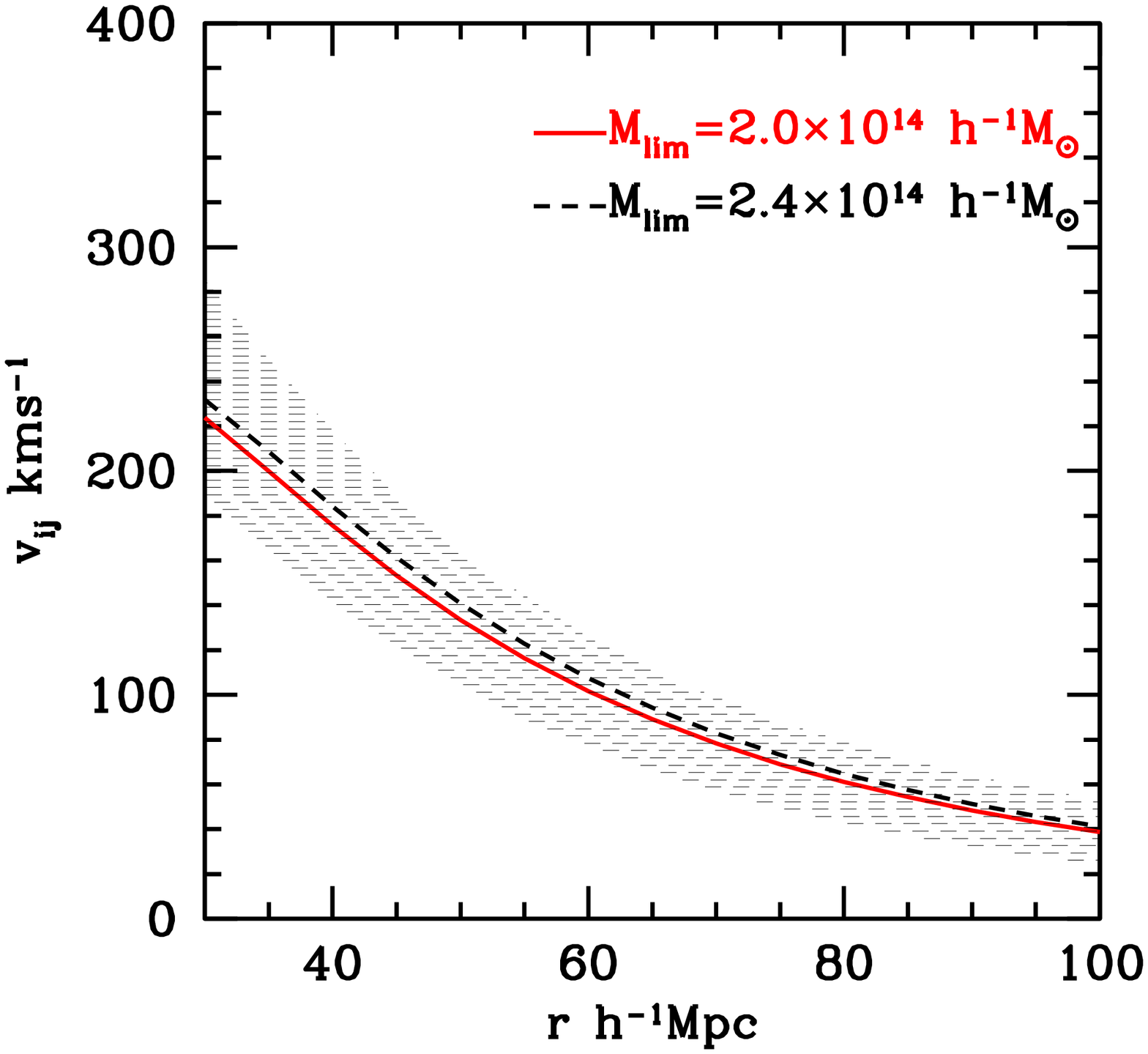}} 
      \resizebox{85mm}{!}{\includegraphics{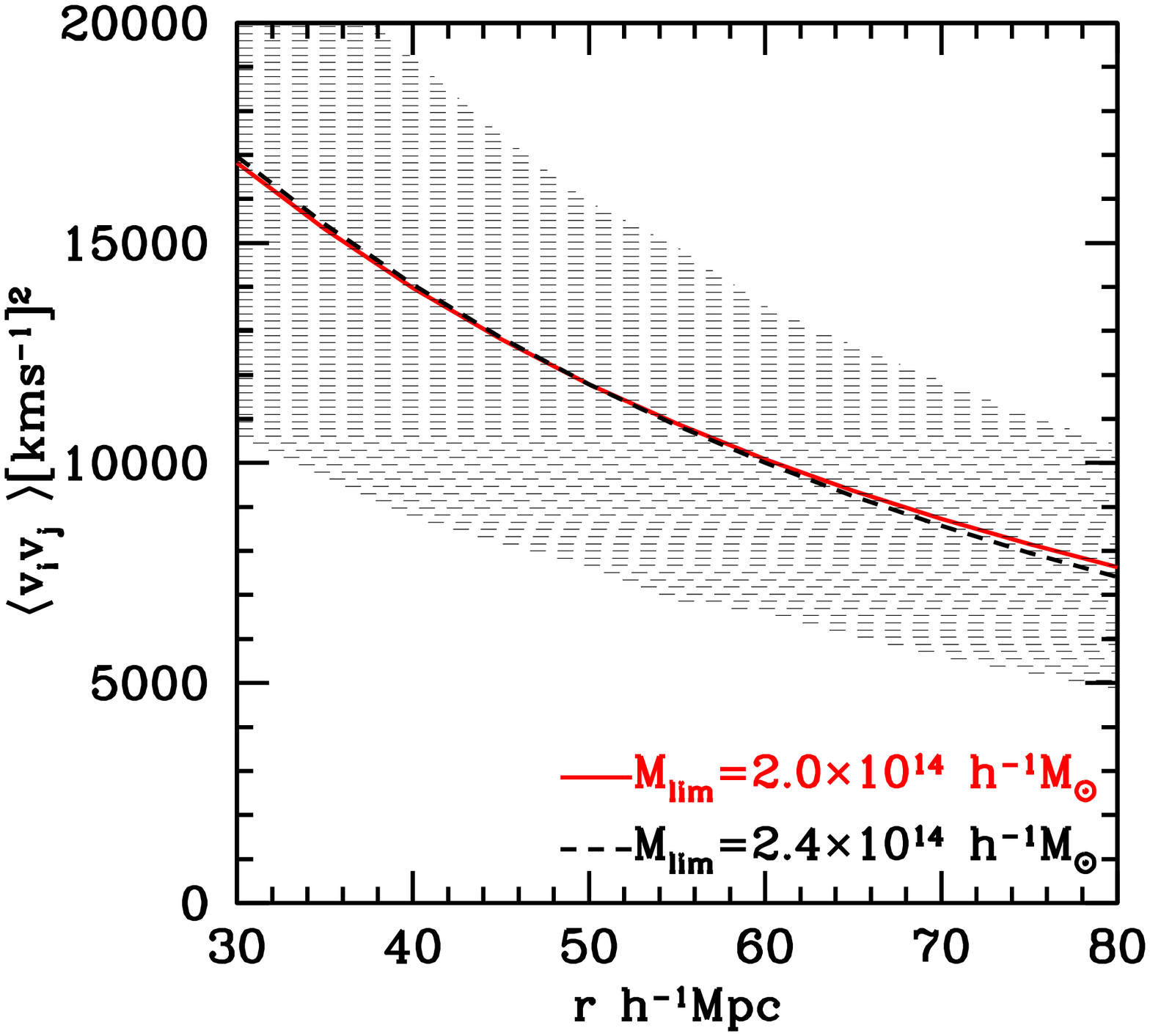}}
    \end{tabular}
    \caption{Velocity statistics for all clusters larger than a minimum mass $M_{\rm min}$, evaluated
for two values of $M_{\rm min}$ differing by 20\%.  (a) The mean pairwise streaming velocity and (b) 
    the perpendicular velocity correlation function.}
    \label{massselection}
  \end{center}
\end{figure*}

To approximate the effect of cluster mass mis-estimates, assume that
for a given sample of galaxy clusters detected via the SZ effect, all of the inferred masses are
off by 20\%. We compute the velocity statistics for clusters with both the actual mass cutoff
and the inferred one using the halo model, and find that this large change in mass selection has minimal
effect: a 20\% offset in minimum mass estimate gives only 2\% to 4\% change for both
velocity statistics. Fig.~\ref{massselection} displays the difference. Therefore, even if the
mass determination of clusters is uncertain at this level, it will result in only small changes
in the underlying cosmological models selected by the data. We emphasize that this is
in marked contrast to the case for cluster number counts.
Although self-calibration techniques provide a possible remedy to the cluster 
count mass-selection bias \citep{mm03,lima05,lima04}, it requires at minimum
a determination of the scatter and bias in cluster photometric 
redshifts to better than 0.03 and 0.003 respectively in order for self-calibration to work \cite{lima07}.
The evolution of cluster properties with redshift also must be of an assumed form.

\section{Systematic Errors from Velocity Misestimates}
\label{sec:xray}

Aside from errors in the inferred cluster mass, which the previous section shows has
little effect on cluster velocity statistics, the cluster velocities themselves are also
subject to systematic errors.
The state of some small volume of gas is characterized by its temperature, density, and bulk
velocity. Measurements of the SZ distortions of the radiation passing through this gas
at three frequencies often have a physical degeneracy which enable measurement of
only two of these gas quantities \cite{aghanim03,holder04,sehgal05}; in particular, this is true for
the ACT frequency bands at 145, 220, and 280 GHz. 
To get the third (typically the gas velocity, the quantity
of interest here), additional information must be obtained. The most convenient source
is a direct determination of gas temperature, either through X-ray observations or through
a theoretical correlation of gas temperature and total thermal SZ distortion. 
In addition, extracting the SZ signal accurately in the presence of foreground emission,
particularly from infrared point sources \cite{borys03,coppin06,kscott08}, can also lead to systematic
errors in inferred velocities if the point sources are not adequately characterized. 

Typically the X-ray temperature $T_X$ of galaxy clusters differs appreciably from the electron gas 
temperature $T_e$.
Numerical simulations \citep{diaferio05,hansen04} 
indicates that using  $T_X$ as a proxy for 
$T_e$ leads to over-estimation of peculiar velocities inferred from SZ measurements
by 10 to 40\%.
This bias is because the estimate of peculiar velocity gets weighted by the ratio of pressure-weighted 
temperature to mass-weighted temperature \cite{knox04}. 
In order to quantify the effect of the difference between $T_X$ and $T_e$ in the estimation of peculiar 
velocity, we use the relation 
$\langle v \rangle_{\theta_b} \propto \langle T_e \rangle_{\theta_b}$
for an unresolved cluster \cite{diaferio05}.
The quantities $\langle..\rangle$ indicates the average quantities over the beam size $\theta_b$;
for brevity we drop the average symbols and assume beam-averaged quantities in this Section. 
We assume the X-ray temperature $T_X$ and the electron temperature $T_e$ are simply related 
through a linear relation $T_e\equiv a + b T_X$.
Numerical studies find $(a,b)=(0.17\,{\rm keV}, 0.69)$ when averaged within the virial radius for a cluster
and $(0.18\, {\rm keV}, 0.53)$ when averaged within three times the virial radius \cite{diaferio05}. 
The relation between the velocity derived from the X-ray temperature 
and from the electron temperature is just 
\begin{equation}
v_{\rm true} = (a / T_X + b)v_{\rm obs},
\label{vpec}
\end{equation}
where $v_{\rm true}$ is the velocity inferred from the electron temperature and $v_{\rm obs}$
is the velocity inferred from the X-ray temperature; Sunyaev-Zeldovich cluster velocity
measurements will be correct when using the electron temperature. 
For a cluster of temperature $T_X=3$ keV, the first term is around 10\% of the second term, and the 
relative contribution decreases further for more massive clusters; we thus neglect the first term,
leaving the true velocity proportional to the observed velocity, 
\begin{equation}
v_{\rm obs} = \beta v_{\rm true}
\end{equation}
with $\beta\equiv 1/b$. 
For the given values of $b=0.69$ and 0.53, the cluster velocity derived using a measured X-ray 
temperature
will be 1.4 and 1.9 times larger than the true velocity derived from the gas temperature. To the
extent that we will not know perfectly the $T_X$--$T_e$ relation, our inferred cluster
velocities will be dominated by our fractional misestimate of $\beta$: a 10\% overestimate of $\beta$
gives about a 10\% overestimate of the cluster velocity.  Note that for a sample of clusters,
$b$ (and thus $\beta$) will likely be easier to infer than $a$ 
since it represents the slope of the $T_e$--$T_X$ relation,
rather than an extrapolation of this relation to $T_X=0$.

Point sources can be modeled in the same way; Ref.~\cite{aghanim04} discusses the
fact that systematic errors in velocity from point sources can be significant even though their contribution
to statistical error may be small. We again assume an observed velocity proportional
to the true velocity; we also add a constant offset, $v_{\rm true} = \beta v_{\rm obs} - v_{\rm off}$. 
Numerical simulations with large numbers of clusters suggest the value $\beta=2$, perhaps
slightly larger than that expected from X-ray temperatures.

To model these effects, we consider a hypothetical cluster velocity survey with
velocities measured for all clusters of mass $M > 2 \times 10^{14} h^{-1}M_{\odot}$ with 
a statistical error of $\sigma_v=300$ km/s,
plus cosmic variance and Poisson noise for a given survey area. This is a rather 
optimistic scenario in terms of the measurement errors. 
We  are interested in quantifying the degradation in dark energy constraints due to systematic errors,
compared to the case of only statistical errors. 
The most optimistic statistical error gives an estimate
with maximum relative systematic error impact; 
hence, the results here represents the worst case scenario.
We also assume a rather high value for $\sigma_8=0.9$. For a lower value $\sigma_8=0.75$, the 
total number of clusters above a given mass threshold will be
reduced by 40\%, which in turn will increase the statistical errors accordingly.

\begin{figure*}
  \begin{center}
    \begin{tabular}{cc}
      \resizebox{85mm}{!}{\includegraphics{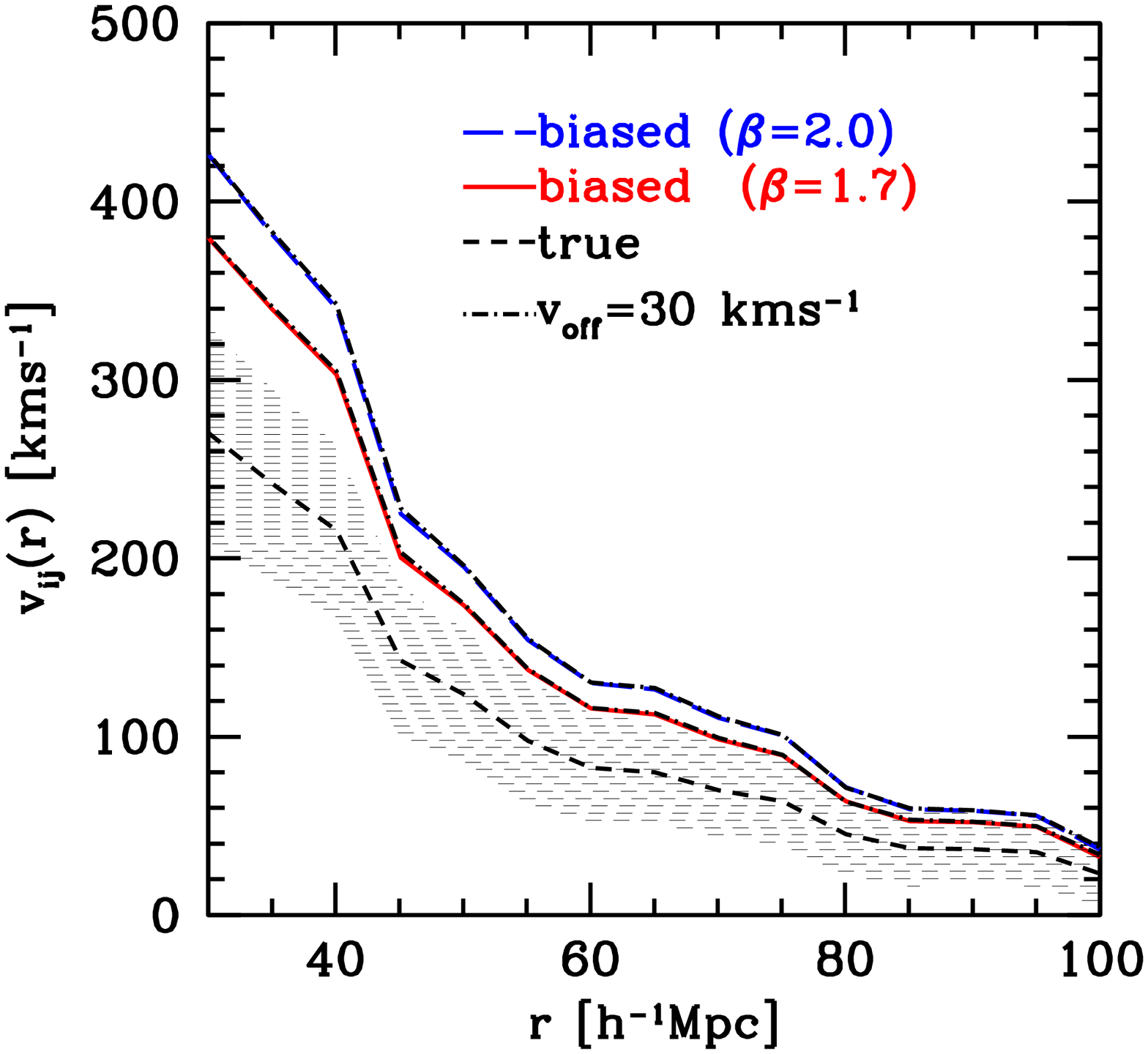}}
\resizebox{85mm}{!}{\includegraphics{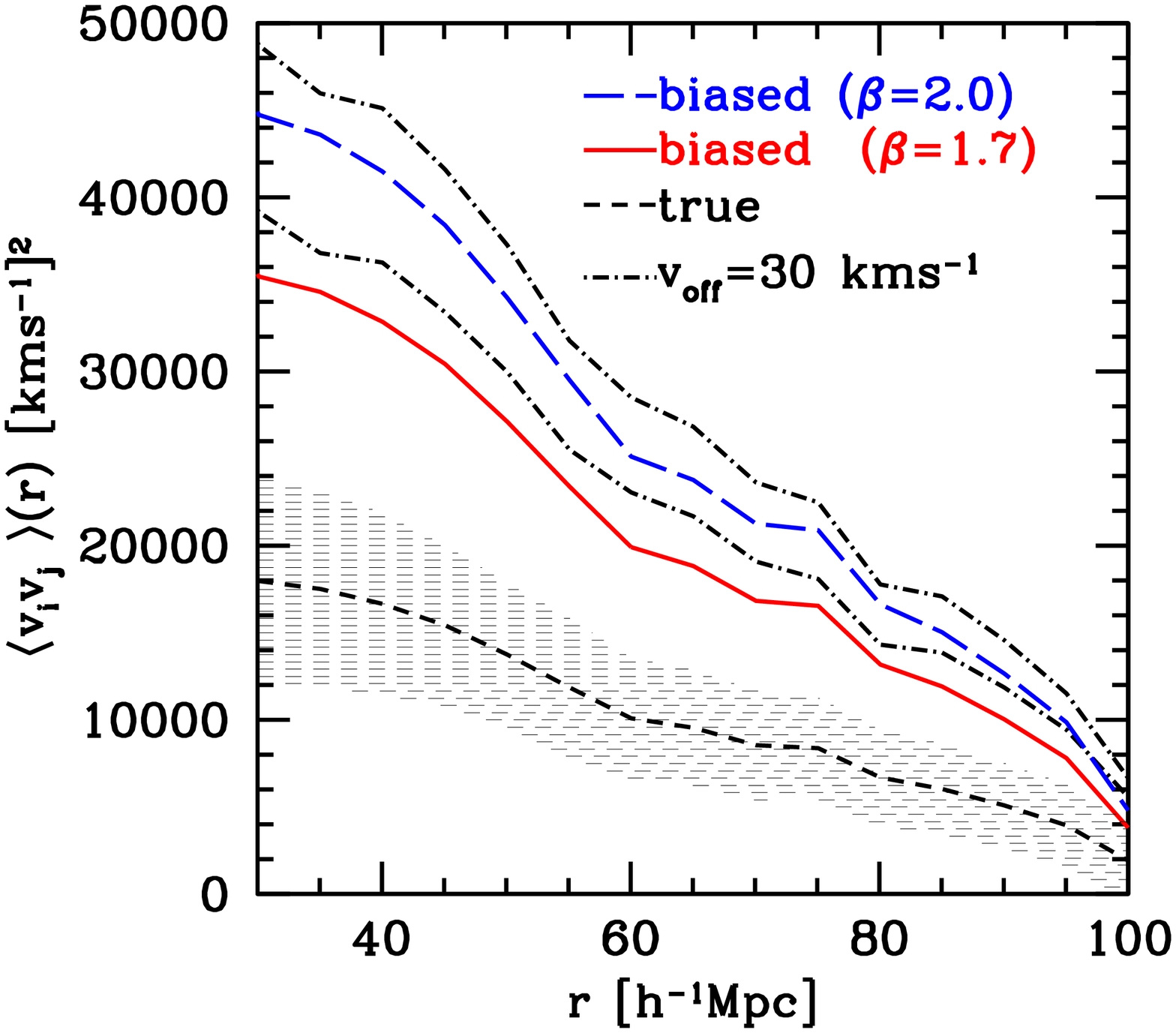}}
    \end{tabular}
    \caption{The effect of systematic errors for the mean pairwise streaming velocity (left) and the velocity 
    correlation function (right). The shaded region with the dashed line shows each statistic obtained from 
    the Virgo dark-matter simulation. The effect of unmodeled systematic
    bias between estimated and actual gas temperature are shown with the red solid line ($\beta=1.7$)
and the blue dashed line ($\beta=2$). Also shown as dot-dashed black lines are the effects of a 
    constant velocity offset $v_{\rm off}=30$ km/sec. Note that $v_{ij}$ has the advantage of being 
    insensitive to velocity offsets.}
    \label{bias}
  \end{center}
\end{figure*}

Fig.~\ref{bias} shows the effect of systematic errors on the pairwise mean streaming velocity 
and the perpendicular velocity correlation function, as determined from the lightcone output
of the VIRGO simulation \cite{virgo}. 
The dashed line surrounded by the shaded region shows the actual value of
the statistics as drawn from the simulation, with inferred statistical errors
assuming a velocity error of  for a 4000 deg$^2$ sky area; see Ref.~\cite{bk07}
for details. The higher offset solid lines show the same quantities except with the individual
cluster velocities biased using $\beta=1.7$ and $\beta=2$, while the dot-dash line shows a constant
velocity offset corresponding to $v_{\rm off}=30$ km/s. Note that a constant offset has
no effect on the mean pairwise streaming velocity, and a relatively small effect on
the correlation function compared to the shift due to $\beta$. Constant velocity offsets
would also be evident in the velocity distribution function, since the entire cluster peculiar veloctiy
sample should have zero mean \cite{bk07}. For the mean pairwise velocity,
a bias corresponding to $\beta=1.7$ shifts the velocity statistic by about 1$\sigma$ statistical
error, while the effect is substantially larger in the velocity correlation function.

In the case of X-ray temperature, we already have reasonable estimates
of the difference in X-ray and electron temperatures, from both analytic and numeric
calculations; the actual bias in velocity statistics will be due only to our error in
understanding this relation, which should be much smaller than the size of the effect
displayed in Fig.~\ref{bias}. The extent to which we can characterize and understand
the effect of the point source population is currently under investigation and requires
a better observational characterization of the relevant sources and their correlation
with galaxy clusters. Ultimately the point sources can be spatially resolved by
observations at submillimeter wavelengths, but doing this over a survey region of several hundred 
square degrees is likely impractical in the foreseeable future. 

The numbers presented here are
a worst-case scenario, should we have a gross misunderstanding of cluster physics,
or completely fail to recognize a substantial source of systematic error in cluster velocity estimates. 
We have simply assumed that we do not account for systematic offsets in X-ray temperature
compared to electron temperature, or systematic errors in cluster velocity estimates due to
point source contamination. We already have detailed estimates of the former, based on
simulations, and the latter is under active study \cite{lin07,righi08,kscott08,wilson08}.
We can also hope to
measure these systematic effects directly from the cluster velocity data; such ``self-calibration''
will be considered below. 

\section{Bias in Dark Energy Parameters}
\label{sec:bias}

In order to study the bias induced in dark energy parameters from systematic errors in the velocity 
statistics, we consider a fiducial cosmology described by the set of cosmological parameters ${\bf p}$ on 
which the velocity field 
depends: the normalization of the matter power spectrum $\sigma_8$, the power law index of the 
primordial power 
spectrum $n_S$, and the Hubble parameter $h$, plus  
the dark energy parameters namely, $\Omega_\Lambda$
and two parameters $w_0$ and $w_a$ describing the redshift evolution of its equation
of state $w(a) = w_0 + (1-a)w_a$. We assume Gaussian priors with variances of $\Delta\sigma_8=0.09$, 
$\Delta n_S=0.015$ \cite{WMAP3} and $\Delta h=0.08$ \cite{hst}.
We then perform a simple Fisher matrix analysis to find the bias on these parameters
from measurements of our two velocity statistics with small systematic errors.

The Fisher information matrix for each of the two statistics is \cite{bk06}
\begin{equation}
F_{\alpha\beta}= \sum_{i,j}\frac{\partial \phi(i)}{\partial p_\alpha}[C^{\phi}_t(ij)]^{-1}\frac{\partial \phi(j)}
{\partial p_\beta}
\label{fisher}
\end{equation}
where $\phi$ stands for either  $v_{ij}(r,z)$ or $\langle v_iv_j\rangle(r,z)$,
$C^{\phi}(ij)$ is the total covariance matrix in each bin. A detail description of the statistics and its 
covariance calculation are given in Ref.~\cite{bk07}.
Assuming the systematic offsets in the velocity statistics are small so 
that the gaussian assumption is valid, the bias in parameter $p$ can be written as \cite{zentner07}
\begin{equation}
\delta p_{\alpha}= \sum_{\beta}[F^{-1}]_{\alpha\beta}\sum_{i,j}\phi(i)_{\rm sys}[C^{\phi}_t(ij)]^{-1}\frac
{\partial \phi(j)}{\partial p_\beta}
\label{fisherbias}
\end{equation}
where $\phi_{\rm sys}=\delta \phi$ the difference between the biased and the true value. Note that  the 
assumption of small offsets may not be valid in our case; nevertheless it gives us an estimate of the 
magnitude of the bias.

Tables~\ref{biasvij} \& \ref{biasvivj} show the dark energy parameter biases for each of the two 
velocity statistics, and for each of two survey areas.
Assuming a measurement error normally distributed with $\sigma_v=300$ km/s, both $v_{ij}$ and 
$\langle v_iv_j \rangle$ put tight constraints on dark 
the energy density   and relatively weak constraints on its equation of state in the absence 
of any systematic bias.
The systematic bias for $w_0$ and $w_a$ is only marginally greater than the no-bias statistical error for 
$\beta=2$, for both the statistics and the survey areas considered; the only exception is for the case 
of $v_{ij}$ for 
a 4000 deg$^2$ survey, where the systematic bias on $w_a$ is 3 times greater than the statistical error.
For $\Omega_{\Lambda}$ the bias is substantial for all survey 
areas and both the statistics;  $v_{ij}$ generally gives a smaller bias on $\Omega_{\Lambda}$
than $\langle v_iv_j  \rangle$. While $\Omega_\Lambda$ is strongly constrained by other
measurements, one virtue of a velocity survey is a completely independent constraint on
$\Omega_\Lambda$. Introducing prior cosmological constraints consistent with projections
for the Planck satellite, velocity statistics will also provide competitive constraints on
$w_0$ and $w_a$ \cite{bk07}. Hence it is important to determine whether self-calibration
of unknown systematic errors will help reduce the bias in determining these parameters. 

We note in passing that non-trivial covariances exist between the various parameters; these
will be considered in greater detail elsewhere using Monte Carlo explorations of the parameter
space.

\begin{table}
\begin{center}
\begin{tabular}{ccccccccc}
\hline
Survey Area & Parameter (p) &$1\sigma$ error ($\Delta p$)& bias ($\delta p$) & $\delta p/\Delta p$&bias 
($\delta p$) & $\delta p/\Delta p$&bias ($\delta p$) & $\delta p/\Delta p$ \\
deg$^2$& & &$\beta=1.4$&$\beta=1.4$&$\beta=1.7$&$\beta=1.7$&$\beta=2.0$&$\beta=2.0$\\\hline
& $\Omega_{\Lambda}\,[0.7]$& 0.03 & 0.04 & 1.46 & 0.075 & 2.56 & 0.11 & 3.7\\
4000 & $w_0\,[-1]$             &0.35 & 0.15 & 0.43 & 0.35 & 0.75 & 0.37 & 1.1\\
& $w_a\,[0]$            &0.56 & 0.65 & 1.15 & 1.13 & 2.0 & 1.6 & 2.9\\\hline
& $\Omega_{\Lambda}\,[0.7]$&0.034&    0.07&      2.0&     0.12&      3.5&     0.17& 5.1\\
 400 &$w_0\,[-1]$&    0.84&    0.07&    0.08&     0.12&     0.14&     0.17&     0.2\\
& $w_a\,[0]$&      1.5&     0.7&     0.5&      1.24&     0.8&      1.8&      1.2\\\hline
\end{tabular}
\end{center}
\caption{The statistical errors $\Delta p$ in dark energy parameters $\Omega_\Lambda$, $w_0$, and 
$w_a$,
and the bias $\delta p$ in these parameters due to systematic error in cluster velocity estimates, 
using the
mean pairwise streaming velocity $v_{ij}(r)$. The fiducial cosmological model has $n_s=1$, 
$\sigma_8=0.9$, $h=0.7$, $\Omega_\Lambda=0.7$, $w_0=-1$, and $w_a=0$, with prior normal errors 
of $\Delta n_s= 0.015$, $\Delta \sigma_8= 0.09$ and $\Delta h= 0.08$ and a spatially flat universe 
assumed.  No priors on dark energy parameters are included.  Cluster velocity normal errors of 
$\sigma_v=300$ km/s are assumed.}
\label{biasvij}
\end{table}

\begin{table}
\begin{center}
\begin{tabular}{ccccccccc}
\hline
Survey Area & Parameter (p) &$1\sigma$ error ($\Delta p$)& bias ($\delta p$) & $\delta p/\Delta p$&bias
($\delta p$) & $\delta p/\Delta p$&bias ($\delta p$) & $\delta p/\Delta p$ \\
deg$^2$& & &$\beta=1.4$&$\beta=1.4$&$\beta=1.7$&$\beta=1.7$&$\beta=2.0$&$\beta=2.0$\\\hline
& $\Omega_{\Lambda}\,[0.7]$ & 0.038 & 0.22 & 5.7 & 0.43 & 11.5 & 0.68 & 17.8\\
4000 & $w_0\,[-1]$             &0.41 & 0.09 & 0.22 & 0.17 & 0.43 & 0.28 & 0.69\\
& $w_a\,[0]$             &0.71 & 0.31 & 0.43 & 0.6 & 0.85 & 0.96 & 1.35\\\hline
& $\Omega_{\Lambda}\,[0.7]$& 0.08&     0.2&      2.7&     0.4&      5.3&     0.7&      8.4\\
400 &  $w_0\,[-1]$ &    0.96&     0.11&     0.12&     0.22&     0.23&     0.35&     0.36\\
& $w_a\,[0]$ &   1.9&     0.5&     0.27&      1.01&     0.54&      1.6&     0.86\\\hline
\end{tabular}
\end{center}
\caption{The same as in Table~\ref{biasvij}, but for the velocity correlation function.} 
\label{biasvivj}
\end{table}

\section{Self-Calibration of Systematic Observables}
\label{sec:selfcal}

One potential method for dealing with systematic errors is to adopt some reasonable
parameterized model for the errors, then solve for these systematic error parameters
along with the cosmological parameters of interest, given the data in hand. 
In this section we study the self-calibration of the variable that models systematic velocity errors 
due to X-ray temperature offset or imperfect point-source subtraction. To this end, we allow $\beta$ 
to co-vary with the cosmological parameters. To allow for redshift evolution of $\beta$, 
we write $\beta=\beta_0/(1+z)^{\gamma}$ with $\beta_0$ being the value of $\beta$ at $z=0$. 
We choose a fiducial value of $\gamma=0$, while fiducial values of $\beta_0=1.4$, 1.7, and 2
correspond to the three values of $\beta$ considered in Sec.~\ref{sec:bias}. We also assume a mild 
normal distribution prior on these parameters with a variance of 50\% . We envisage such moderate
priors can be obtained using numerical simulation studies and follow-up observations.  
We perform a fisher matrix analysis using Eq.~(\ref{fisher}) with the cosmological parameters plus the 
two systematic 
parameters $\beta_0$ and $\gamma$, then marginalize over $\beta_0$ and $\gamma$
to get the dark energy parameter constraints. Cosmological parameters will necessarily
have their constrains weakened, but if the model for the systematic errors is an accurate 
representation of the actual systematic errors, the bias in cosmological parameters will be
reduced. 

Results are shown in Table~\ref{calvij} \& \ref{calvivj} for the two 
statistics $v_{ij}$ and $\langle v_iv_j \rangle$ and two survey 
areas. For 
$v_{ij}$, the degradation in the constraint on $\Omega_{\Lambda}$ varies from 3\% to 10\% for 
$\beta=1.4$ to 2.0. For $w_0$, the degradation varies from 
6\% to 18\% and for $w_a$ it is 27\% to 40\% for a 4000 deg$^2$ survey area. For 400 deg$^2$ the 
relative degradation of the  constraints is smaller since the statistical error is larger than for greater sky
area.  For $\langle v_iv_j \rangle$ the degradation is larger 
than for $v_{ij}$ since it varies as $\beta^2$. 
Table~\ref{syspar} gives the constraints on the parameters $\beta_0$ and $\gamma$ that are used to 
describe the systematic velocity errors: $v_{ij}$ gives 20\% and 30\% constraints on $\beta_0$ and 20\% 
and 40\%
constraint on $\gamma$ for the two survey areas 4000 and 400 deg$^2$ respectively.

\begin{table}
\begin{center}
\begin{tabular}{ccccccccc}
\hline
Survey Area & Parameter (p) &$1\sigma$ error ($\Delta p$)& self-cal($\Delta_s p$) & ${\Delta_s p}/
{\Delta p}$&self-cal($\Delta_s p$) & ${\Delta_s p}/{\Delta p}$&self-cal($\Delta_s p$) & ${\Delta_s p}/
{\Delta p}$ \\
deg$^2$& & &$\beta=1.4$&$\beta=1.4$&$\beta=1.7$&$\beta=1.7$&$\beta=2.0$&$\beta=2.0$\\\hline
&$\Omega_{\Lambda}\,[0.7]$ &0.029&    0.03&      1.03&    0.031&      1.07&
    0.032&      1.10\\
 4000 & $w_0\,[-1]$ &   0.35&     0.37&      1.06&     0.38&      1.08&
     0.41&      1.175\\
  & $w_a\,[0]$&  0.56   &0.71&1.27&      0.75&        1.35& 0.78&1.4\\\hline
& $\Omega_{\Lambda}\,[0.7]$ &0.034&    0.036&    1.06&    0.036&   1.06&
    0.037&     1.07\\
400 & $w_0\,[-1]$ &   0.84&    0.87&      1.04&     0.89&      1.06&
     0.93& 1.11\\ 
&  $w_a\,[0]$ &   1.5&      1.7&      1.13& 1.9    &      1.27&
      2.2&      1.46\\\hline
\end{tabular}
\end{center}
\caption{Constraints with the self-calibration of the systematic parameters $\beta_0$ and $\gamma$ 
for $v_{ij}$. A prior of $\pm 0.5$ is assumed on the systematic parameters namely $\beta_0$ \& $
\gamma$. Note that $\Delta_s p$ \& $\Delta p$ denotes the $1-\sigma$ statistical error on dark energy 
parameters when nuisance parameters are included and not included respectively}
\label{calvij}
\end{table}

\begin{table}
\begin{center}
\begin{tabular}{ccccccccc}
\hline
Survey Area & Parameter (p) &$1\sigma$ error ($\Delta p$)& self-cal($\Delta_s p$) & ${\Delta_s p}/
{\Delta p}$&self-cal($\Delta_s p$) & ${\Delta_s p}/{\Delta p}$&self-cal($\Delta_s p$) & ${\Delta_s p}/
{\Delta p}$ \\
deg$^2$ & & &$\beta=1.4$&$\beta=1.4$&$\beta=1.7$&$\beta=1.7$&$\beta=2.0$&$\beta=2.0$\\\hline
& $\Omega_{\Lambda}\,[0.7]$& 0.038&    0.042&      1.09&    0.044&      1.15&
    0.047&      1.23\\
4000 & $w_0\,[-1]$ &   0.41&     0.65&      1.6&     0.8&      1.97&
     1.0&      2.4\\
& $w_a\,[0]$   &  0.71&      2.64&      3.7&      3.1&      4.4&
      3.6&      5.2\\\hline
& $\Omega_{\Lambda}\,[0.7]$&  0.08&    0.088&      1.09&    0.095&      1.18&
     0.11&      1.31\\
400 &  $w_0\,[-1]$  &  0.96&      1.6&      1.69&      1.97&      2.0&
      2.42&      2.52\\
& $w_a\,[0]$   &  1.9&      5.3&      2.8&      6.7&      3.6&
      8.3&      4.4\\\hline
\end{tabular}
\end{center}
\caption{Same as Table~\ref{calvij} but for $\langle v_iv_j \rangle$.}
\label{calvivj}
\end{table}

\begin{table}
\begin{center}
\begin{tabular}{ccccccc}
\hline
Survey Area& & $v_{ij}$ & & &$\langle v_iv_j \rangle$& \\
deg$^2$ &$\beta=1.4$&$\beta=1.7$&$\beta=2.0$&$\beta=1.4$&$\beta=1.7$&$\beta=2.0$\\\hline
4000 & $0.24\,[0.25]$ & $0.21\,[0.23]$ & $0.18\,[0.22]$ & $0.09\,[0.2]$ & $0.085\,[0.16]$ & $0.08\,[0.13]$\\
400 & $0.32\,[0.42]$&$0.3\,[0.4]$ & $0.27\,[0.39]$ & $0.16\,[0.38]$ & $0.17\,[0.33]$ & $0.17\,[0.29]$\\\hline
\end{tabular}
\end{center}
\caption{Constraints on the parameters $\beta_0$ and $\gamma$ used to model the systematic offset.
Constraints are shown as $\Delta \beta_0\,[\Delta\gamma]$.}
\label{syspar}
\end{table}

\section{Summary and Context}

Galaxy cluster velocity surveys, with velocities determined from kinematic Sunyaev-Zeldovich
measurements, have the potential to provide constraints on dark energy parameters competitive with 
other methods. Their challenge is attaining sufficient sensitivity over a large enough sky region; the 
kinematic Sunyaev-Zeldovich signal is typically only a tenth of the thermal SZ signal
for the same galaxy cluster. But while cluster number counts based on the thermal SZ signal
probe dark energy with a signal that is easier to detect, cluster velocities have the tangible
advantage that they are nearly insensitive to the uncertainties in the connection between the
underlying cluster mass and the observed cluster SZ signal that will substantially 
impact cluster number counts. We have demonstrated
this fact explicitly by computing the cluster mean streaming velocity and velocity
correlation function from the halo model, finding that both statistics
are nearly independent of the cluster mass threshold chosen. Galaxy clusters serve
as tracers of the underlying large-scale velocity field, independent of their masses. 

Cluster velocities are not without their own systematic errors, however; in particular,
the velocities inferred from kSZ measurements may have systematic errors arising from
a mis-estimated relation between the cluster X-ray temperature and its gas temperature,
or from an incorrect correlation between the thermal SZ signal and the cluster gas temperature. Either
might be used to extract the cluster velocity from three-band microwave measurements. 
Further complications arise from infrared point sources, which have variable spectral
indices and can be substantially correlated with galaxy cluster positions. We have considered
a simple toy model which assumes a linear relationship between the measured
and actual cluster velocity. If this relation is not well understood, we have quantified the bias
it can induce in inferred values of cosmological parameters. This bias can be significant
compared to the corresponding statistical errors, although reasonable levels of
understanding of these correlations will likely reduce the systematic errors to below
the level of statistical errors.






This work has considered the cluster peculiar velocity to be the basic observable.
Velocity statistics are relatively straightforward to compute in particular cosmological models
\cite{sheth01, sheth02, sheth04, bk07}, but the kinematic SZ effect actually measures a line-of-sight
integral of
the gas density times the peculiar velocity, proportional to the cluster gas momentum. 
The systematic errors associated with inferring the cluster velocity, particularly
the discrepancy between X-ray temperature and gas temperature, can be avoided
completely by measuring the cluster {\it gas momentum} via the kinematic SZ distortion,
then comparing these to theoretical distributions of cluster momenta instead of
velocities. The difficulty then becomes accurate prediction of cluster momentum
statistics, which must be done from simulations. While total cluster momentum
is relatively straightforward to extract from large-volume cosmological dark
matter simulations, the cluster gas momentum depends on the cluster gas
fraction. This fraction is expected on general grounds to be relatively constant
for large clusters, since it is difficult for gas to escape from their deep gravitational
wells, but their star formation history might vary significantly and depend in detail on
the physics of star formation and energy feedback from stars and active galaxies. This question is
addressed for lower-mass galaxy groups in \cite{quasar07}, but realistic estimates for
galaxy clusters await larger-volume cosmological hydrodynamic simulations. 
Uncertainty in cluster baryon fraction could be a significant systematic error
when using cluster momentum statistics to constrain cosmology.


To date, a galaxy cluster peculiar velocity has never been measured directly: no
Sunyaev-Zeldovich measurements have had the combination of arcminute angular
resolution and micro-Kelvin sensitivity needed to make a statistically significant 
detection of the kinematic SZ signal. (Velocity upper limits have been obtained for a few clusters \cite{benson03,holz97}.) 
This situation will change in the near future; both ACT \cite{kosowsky06,fowler06}
and SPT \cite{ruhl05} likely will have sufficient angular resolution, sensitivity, and frequency coverge,
and are currently taking data. The first direct measurement of the peculiar velocities of objects at 
cosmological
distances will be a signature achievement, and when extended to large
surveys will give an independent route to constraining dark energy properties \cite{bk07}
and properties of gravitation 
(S.~Bhattacharya \& A.~Kosowsky, in preparation (2008)).  
Capitalizing on this opportunity requires sufficient
control over systematic errors, as with all dark energy probes. As we have shown here,
cluster velocities have a large advantage over cluster number counts because the selection
function of the clusters has little impact on the cluster statistics we want to measure; cluster number
counts, in contrast, are equally sensitive to the selection function as to cosmology.  Systematic
errors in velocity estimates for individual clusters can potentially cause significant bias
in the inferred values of dark energy parameters, but based on the calculations presented here
we anticipate that careful observations
and modeling of galaxy clusters can limit these systematics to levels below the statistical
errors in cosmological constraints.

\section*{Acknowledgement}

This work has been supported by NSF grant AST-0408698 to the ACT project, and
by NSF grant AST-0546035. SB has been partially supported by a Mellon Graduate Fellowship at the
University of Pittsburgh.

\bibliography{paper3v3PRD}

\end{document}